\documentclass[10pt,aps,prl, twocolumn,superscriptaddress,floatfix]{revtex4}
\usepackage{graphicx}
\usepackage[english]{babel}
\usepackage{amsmath}
\usepackage{amssymb}
\usepackage[usenames,dvipsnames]{color}
\usepackage{color}
\definecolor{Blue}{rgb}{0.3,0.3,0.9}
\definecolor{orange}{rgb}{1,0.5,0}
\usepackage{subfigure}
\usepackage{bm}
\usepackage{titlesec}
\usepackage{mathrsfs}
\usepackage{gensymb}
\usepackage{esint}
\usepackage[unicode=true,  bookmarks=false,  breaklinks=false,pdfborder={0 0  1},backref=false,colorlinks=true]  {hyperref}
\newcommand{\bq}{{\mathbf q}}

\newcommand{\s}{\sigma}

\newcommand{\Sx}{S^x}
\newcommand{\Sy}{S^y}
\newcommand{\Sz}{S^z}

\begin{document}
\author{Mohammad F. Maghrebi}
\email{magrebi@umd.edu}

\affiliation{Joint Quantum Institute, NIST/University of Maryland, College Park, Maryland 20742, USA}

\affiliation{Joint Center for Quantum Information and Computer Science, NIST/University of Maryland, College Park, Maryland 20742, USA}

\author{Zhe-Xuan Gong}

\affiliation{Joint Quantum Institute, NIST/University of Maryland, College Park, Maryland 20742, USA}

\affiliation{Joint Center for Quantum Information and Computer Science, NIST/University of Maryland, College Park, Maryland 20742, USA}

\author{Alexey V. Gorshkov}

\affiliation{Joint Quantum Institute, NIST/University of Maryland, College Park, Maryland 20742, USA}

\affiliation{Joint Center for Quantum Information and Computer Science, NIST/University of Maryland, College Park, Maryland 20742, USA}

\title{Continuous symmetry breaking and a new universality class in 1D long-range interacting quantum systems}
\begin{abstract}
  Continuous symmetry breaking (CSB) in low-dimensional systems, forbidden by the Mermin-Wagner theorem for short-range interactions, may take place in the presence of slowly decaying long-range interactions. Nevertheless, there is no stringent bound on how slowly interactions should decay to give rise to CSB in 1D quantum systems at zero temperature.
  Here, we study a long-range interacting spin chain with $U(1)$ symmetry and power-law interactions $V(r)\sim1/r^\alpha$, directly relevant to ion-trap experiments.
  Using bosonization and renormalization group theory, we find CSB for $\alpha$ smaller than a critical exponent $\alpha_c(\le 3)$ depending on the microscopic parameters of the model.
  Furthermore, the transition from the gapless XY phase to the gapless CSB phase is mediated by the breaking of conformal symmetry due to long-range interactions, and is described by a {\it new} universality class akin to the Berezinskii-Kosterlitz-Thouless transition.
  Our analytical findings are in good agreement with a numerical calculation. Signatures of the CSB phase should be accessible in existing trapped-ion experiments.
\end{abstract}

\maketitle

Long-range interacting systems have recently attracted great interest as they emerge in numerous setups in atomic, molecular, and optical (AMO) physics \cite{saffman10,schauss12,firstenberg13,yan13,aikawa12,lu12,childress06,balasubramanian09,weber10,dolde13,gopalakrishnan11,islam13,britton12,Richerme14,Jurcevic14}.
The advent of  AMO physics has offered the intriguing possibility of
simulating many-body systems which have been extensively studied theoretically in condensed matter physics \cite{Bloch08,Bloch12,Lewenstein12}.
While many properties of long-range interacting systems derive from their short-range counterparts, long-range interactions also give rise to novel phenomena \cite{Kivelson04,Pfau09,Buchler12}.
In particular, they can induce spontaneous symmetry breaking in low-dimensional systems, which, for short-range interactions,  is forbidden by the Mermin-Wagner theorem  \cite{Mermin66}. Such possibilities have been studied at finite temperature \cite{deSousa05,Buchler12,Bruno01}, where stronger versions of the Mermin-Wagner theorem have been proven for long-range interacting spin systems \cite{Bruno01}. On the other hand, this subject is much less investigated at zero temperature \footnote{While, strictly speaking, the Mermin-Wagner theorem forbids CSB in 1D and 2D systems at finite temperature \cite{Mermin66}, it is also  believed to generically forbid CSB in zero-temperature 1D systems.}. As long-range interactions effectively change the dimensionality of the system, the emergence of CSB for sufficiently slowly-decaying interactions is not surprising;
however, the equivalents of the stringent bounds at finite temperature \cite{Bruno01} are not known.
Furthermore, the as yet unexplored quantum phase transition from the CSB phase to other 1D quantum phases is rather exotic since, at zero temperature, the phases separated by this transition typically occur in different dimensions.
Finally, with the recent advances of the ion-trap quantum simulator in tuning the power of long-range interactions \cite{islam13,Richerme14,Jurcevic14},
the above questions appear to be of immediate experimental relevance.

In this paper, we consider the ferromagnetic XXZ spin-1/2 chain with power-law interactions $V(r)\sim 1/r^\alpha$. We find that the continuous $U(1)$ symmetry is spontaneously broken for a sufficiently slow decay of the interaction below a critical value of the exponent $\alpha_c(\le 3)$ that depends on the microscopic parameters of the model;
this has to be contrasted with a simple spin-wave analysis that, ignoring vortices, would give $\alpha_c=3$.
Exploiting a number of analytical techniques such as spin-wave analysis, bosonization, and renormalization group (RG) theory, as well as a numerical density matrix renormalization group (DMRG) analysis, we explore the phase diagram, and identify the phase transitions between different phases. In particular, we find a new universality class describing the phase transition between the CSB and XY phases, similar to, but with important differences from, the Berezinskii-Kosterlitz-Thouless transition. The signatures of the CSB phase should be accessible in existing trapped-ion experiments.

{\it Model.---}Let us consider the long-range interacting XXZ chain
\begin{equation}\label{Eq. XXZ Hamiltonian}
  H=\sum_{i> j}\frac{1}{|i-j|^\alpha}\left(-\Sx_i\Sx_j-\Sy_i\Sy_j+J_z \Sz_i\Sz_j\right),
\end{equation}
with $S^{x,y,z}=\sigma^{x,y,z}/2$ where $\s$s are the Pauli matrices.  Note that $J_z$ can be either positive or negative, while the $S^x$-$S^x$ and $S^y$-$S^y$ interactions are ferromagnetic. This model has a $U(1)$ symmetry with respect to rotations in the $x$-$y$ plane.

To explore the phase diagram at zero temperature, we will exploit field theory techniques and, specifically, use bosonization
\cite{GiamarchiBook,SachdevBook}.
However, with long-range interactions between all pairs of spins, bosonizing the spin Hamiltonian is rather complicated, at least at a quantitative level. Nevertheless, to capture the essential features of the phase diagram, we can split the Hamiltonian into two parts: the short-range part of the Hamiltonian
and the asymptotic long-range interaction terms.

We start with the short-range part of the Hamiltonian. The bosonization technique maps the Hamiltonian in Eq.~(\ref{Eq. XXZ Hamiltonian})
to one in terms of the bosonic variables $\phi$ and $\theta$ defined in the continuum, which satisfy the commutation relation
\begin{equation}\label{Eq. Algebra}
  [\nabla \phi(x), \theta(y)]=i\pi \delta(x-y)\,.
\end{equation}
The short-ranged Hamiltonian maps to the sine-Gordon model \cite{GiamarchiBook},
\begin{align}\label{Eq. Hamiltonian SR}
  H_{\rm SR}=\frac{u}{2\pi} &\int dx \left[\frac{1}{K}(\nabla \phi)^2+K(\nabla \theta)^2\right]\nonumber \\
  -\frac{2 g}{(2\pi a_c)^2} &\int dx \cos\left[4\phi(x)\right],
\end{align}
with $a_c$ a short-wavelength cutoff, $K$ the so-called Luttinger parameter, $u$ a velocity scale, and $g$ the strength of the cosine interaction term; the values of these parameters have been computed in the Supplemental Material \cite{supp} by including terms up to the next-nearest neighbor in Eq.~(\ref{Eq. XXZ Hamiltonian}) perturbatively in $J_z$ and $1/2^\alpha$. Higher-order neighbors modify the parameters in Eq.\ (\ref{Eq. Hamiltonian SR}), and induce higher-order harmonics which can nevertheless be neglected as they are less relevant in the RG sense.

To find the long-range part of the Hamiltonian, we first identify the spin operators in terms of the bosonic fields $\phi$ and $\theta$. Defining the raising and lowering spin operators $S^\pm=(\Sx \pm i \Sy)/2$, one can approximately identify \cite{GiamarchiBook,SachdevBook}
\begin{equation}
  S^\pm_{j}\sim e^{\pm i \theta (x_j)}, \qquad \Sz_j\sim \nabla \phi (x_j)\,,
\end{equation}
where $x_j$ is the position of the spin at  site $j$ \footnote{The oscillating phase factor $(-1)^j$ is absent in the expression for $S^\pm$ due to our choice of (ferromagnetic) sign in the Hamiltonian (\ref{Eq. XXZ Hamiltonian}).}. More generally, the spin operators can be expanded in a series of  harmonics $e^{i p \phi}$; however, we have dropped those with $p\ge 1$ as they give rise to less relevant terms in the RG sense. With the above identification, the long-range $S^z$-$S^z$ interaction takes the form
\(
  \int dx dy \, {|x-y|^{-\alpha}}\nabla \phi(x)\,\nabla \phi(y)\,,
\)
which, in momentum space, is proportional to $|q|^{\alpha+1} |\phi(q)|^2$. We shall restrict ourselves to $\alpha>1$, that is, the exponent is larger than the spatial dimensionality, so that the Hamiltonian (\ref{Eq. XXZ Hamiltonian}) has a well-defined thermodynamic limit. With this assumption, the long-range $S^z$-$S^z$ interaction is irrelevant compared to the gradient term in $\phi$ (proportional to $q^2 |\phi|^2$) in Eq.~(\ref{Eq. Hamiltonian SR}) and can thus be neglected. On the other hand, the long-range $S^x$-$S^x$ and $S^y$-$S^y$ interactions can be cast as
\begin{equation}\label{Eq. Hamiltonian LR}
  H_{\rm LR}= - g_{\rm LR}\int dx dy \, \frac{1}{|x-y|^\alpha} \cos\left[\theta(x)-\theta(y)\right],
\end{equation}
with $g_{\rm LR}$ the strength of long-range interactions.
The total (bosonized) Hamiltonian is the sum of the short- and long-range parts given by Eqs.~(\ref{Eq. Hamiltonian SR}) and (\ref{Eq. Hamiltonian LR}), respectively. The cosine terms in Eqs.\ (\ref{Eq. Hamiltonian SR}) and (\ref{Eq. Hamiltonian LR}) involve non-commuting fields and thus compete with each other. To determine which one dominates, we shall resort to renormalization group theory.

{\it Quantum phases.---}To find the phase diagram, we perform an RG analysis that is perturbative $g$ and $g_{\rm LR}$. The quadratic terms in Eq.\ (\ref{Eq. Hamiltonian SR}) yield the scaling dimensions (characterizing scaling properties under spacetime dilations)
\cite{GiamarchiBook,SachdevBook}
\begin{equation}
  \mbox{dim}\left[e^{ip\phi}\right]=\frac{p^2K}{4}\,,\qquad \mbox{dim}\left[e^{ip\theta}\right]=\frac{p^2}{4K}\,.
\end{equation}
The RG equations for the interaction coefficients $g$ and $g_{\rm LR}$ then read (space-time rescaled by $e^{-dl}$)
\begin{align}\label{Eq. RG for g and gLR}
  \frac{dg}{dl}=\left(2-4K\right)g\,, \quad \frac{dg_{\rm LR}}{dl}=\left[3-\alpha-1/(2K)\right]g_{\rm LR}\,.
\end{align}
Note that the value of $K$ itself also depends on $\alpha$.
In deriving the flow of $g_{\rm LR}$, we have used the fact that $x$ and $y$ in Eq.~(\ref{Eq. Hamiltonian LR}) are far separated.

Equation (\ref{Eq. RG for g and gLR}) gives rise to several phases depending on whether the interaction terms are relevant, and which one is more relevant. When both $g$ and $g_{\rm LR}$ are irrelevant,
the cosine terms can be dropped \footnote{This, of course, is based on the assumption that $g$ and $g_{\rm LR}$ can be treated perturbatively.}. In this case, one finds an XY-like phase known as the Tomonaga-Luttinger (TL) liquid.
In this phase, correlation functions decay algebraically with exponents determined by $K$ \cite{GiamarchiBook}.
Nevertheless, there is no true $U(1)$ symmetry breaking as $\langle S_i^+ S_j^- \rangle \to 0$ for $|i-j| \to \infty$. This phase is described by a conformal field theory with the central charge $c=1$ as long-range interactions are irrelevant. When the local interaction term is relevant, and more relevant than the non-local one,
the latter can be dropped, while the former gaps out the system. This regime corresponds to an Ising phase, which occurs for a sufficiently large $|J_z|$: An antiferromagnetic (AFM) phase emerges for large and  positive, but $\alpha$-dependent, values of $J_z$, while a ferromagnetic (FM) Ising phase appears for all $J_z<-1$ as shown in \cite{supp} via a spin-wave analysis.
We stress that all the above phases also exist in the absence of long-range interactions; the presence of such terms, however, modifies the boundaries between these phases.

\begin{figure}
   \centering
    \hskip -.2in \includegraphics[width=8.5cm]{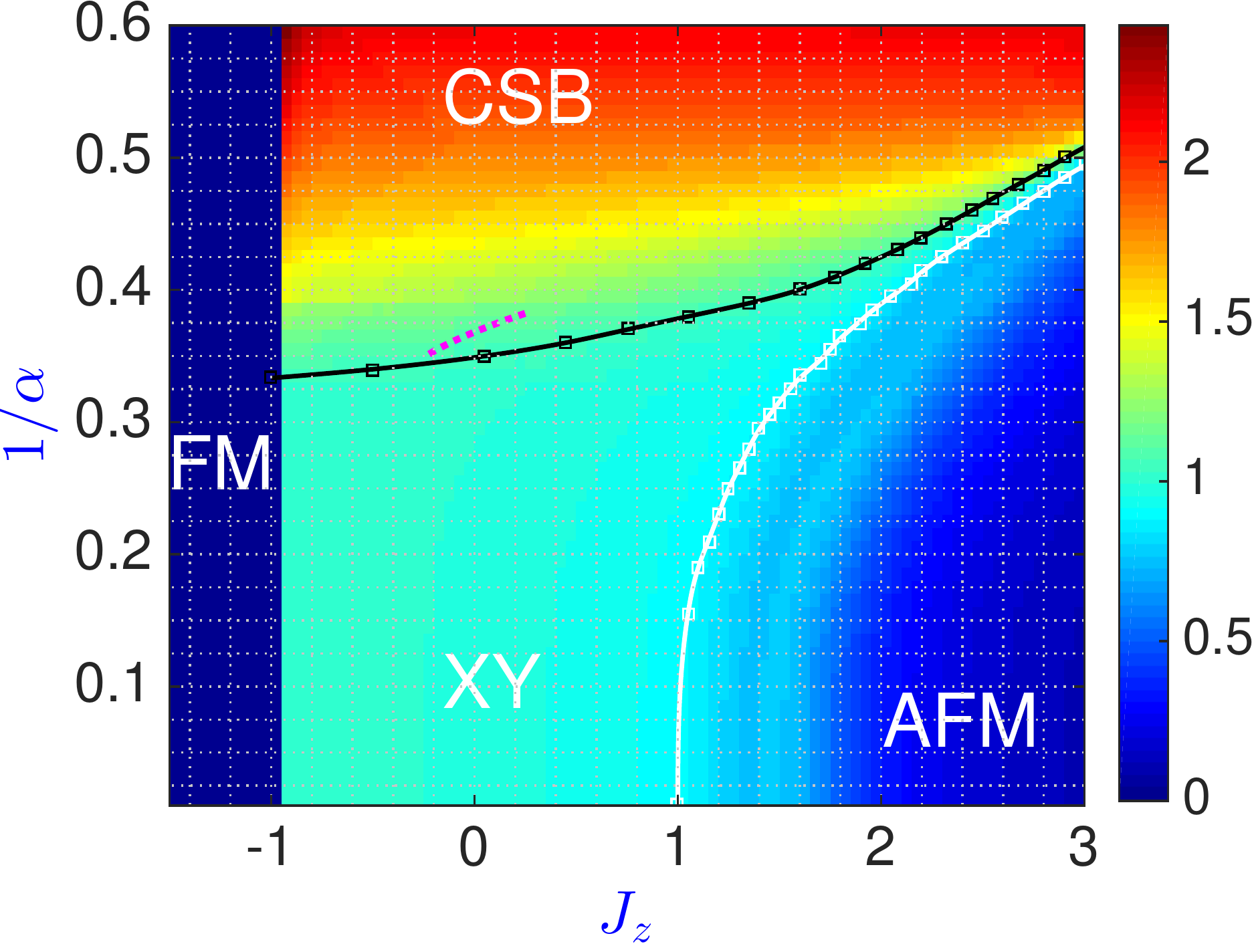}
    \caption{Phase diagram for the Hamiltonian (\ref{Eq. XXZ Hamiltonian}) based on a finite-size DMRG calculation of the effective central charge $c_{\text{eff}}=6[S(N_1)-S(N_2)]/[\log(N_1)-\log(N_2)]$ \cite{calabrese_entanglement_2004}. Here $S(N)$ is the ground-state entanglement entropy for a chain of size $N$ split in two equal halves. We choose $N_1=100$ and $N_2=110$ in our calculation. The XY phase has conformal symmetry and is identified by $c_{\text{eff}}=1$. The XY-to-CSB phase boundary is numerically obtained by finding the place where $c_{\text{eff}}$ starts to increase appreciably ($4\%$) above $1$ (the black squares fitted by the black
line). The dashed (purple) line is the XY-to-CSB transition line obtained from  perturbative field theory calculation in \cite{supp}. The XY-to-AFM phase boundary is obtained by finding the place where $c_{\text{eff}}$ starts to decrease appreciably ($1\%$) below its value at $J_z = 1$ and $\alpha = \infty$ (the white squares fitted by the white line).}
    \label{Fig. SpinHalf}%
\end{figure}

We are mainly interested in a regime where the long-range interaction term is (more) relevant, i.e., $3-\alpha-1/(2K)>0$. Specifically, this implies $\alpha\le3$ as a necessary condition for the long-range interaction to be relevant. In this regime, one can drop the local cosine term, and the model can be described by the Euclidean action
\begin{align}\label{Eq. Action 2}
  I=&\frac{K}{2\pi u} \int \!\!d\tau dx \left[(\partial_\tau \theta)^2+u^2 (\nabla \theta)^2\right]\nonumber\\
  -& g_{\rm LR}\int\!\!\frac{d\tau dx dy }{|x-y|^\alpha} \cos\left[\theta(\tau,x)-\theta(\tau,y)\right]\,,
\end{align}
where the $\nabla \phi$ term in Eq.~(\ref{Eq. Hamiltonian SR}), being conjugate to $\theta$ [Eq.~(\ref{Eq. Algebra})], is replaced by the (imaginary) time derivative $\partial_\tau \theta$ up to a prefactor.
Since $g_{\rm LR}$ grows under RG, the value of the corresponding cosine term is pinned, i.e., $\theta(x)\approx \theta_0 =\mbox{const}$. This, in turn, implies a finite expectation value of the spin in the $x$-$y$ plane,
\(
  \langle S^+_j\rangle \sim e^{i\theta_0}.
\)
It thus appears that the ground state breaks a continuous symmetry. To examine the effect of fluctuations,
we expand the cosine in Eq.~(\ref{Eq. Action 2}) to  quadratic order, and combine it with the quadratic terms in Eq.~(\ref{Eq. Action 2}) to find
\(
I\sim \int d\omega dq \, \left(\omega^2 + q^2+ |q|^{\alpha-1}\right) \, |\theta(\omega,q)|^2,
\)
where we have dropped various coefficients for convenience and taken $\theta_0=0$ without loss of generality.
Clearly, the term proportional to $q^2$ can be dropped compared to $|q|^{\alpha-1}$ for $\alpha<3$; in this case, long-range interactions are dominant and break the conformal symmetry \cite{Pupillo14,vodola15}.
The long-distance correlation of $S^\pm$ is given by
\begin{align}\label{Eq. cs breaking criterion}
  \!\!\!\!\left\langle S^+_iS^-_j \right\rangle \!\sim \!e^{-\left\langle\left[\theta(\tau,x_i)-\theta(\tau,x_j)\right]^2\right\rangle/2}
    =\exp\!\!\left[-{R_{ij}^{-\frac{3-\alpha}{2}}}\right]\!,
\end{align}
where $R_{ij}=|x_i-x_j|$. (We have not kept track of the coefficients in the exponent.)
Notice that $\left\langle S^+_iS^-_j  \right\rangle \not \rightarrow 0$ const as $R_{ij} \rightarrow \infty$.
Therefore, fluctuations respect the continuous symmetry breaking in this phase, in sharp contrast with the destruction of order in short-range interacting systems \cite{Mermin66}. We conclude that CSB may be realized for sufficiently small values of $\alpha(<3)$.
The above findings are consistent with the phase diagram in Fig.~\ref{Fig. SpinHalf} obtained numerically using the finite-size DMRG method \cite{schollwock_densitymatrix_2011,crosswhite_applying_2008,MPSfootnote}.
It is worth pointing out that the quadratic action, after dropping the $q^2$ term, is exact in the RG sense; possible higher-order terms that respect the $U(1)$ symmetry are irrelevant. Specifically, the critical dynamic exponent, determining the relative scaling of space and time coordinates, is given exactly by
\begin{equation}
  z=\frac{\alpha-1}{2}\,\,<1\,.
\end{equation}
The fact that $z<1$ indicates that the `light-cone' characterizing the causal behavior in the CSB phase is sublinear. The response function for this model is studied in great detail in Ref.~\cite{maghrebi15g}, and is shown to take a universal scaling form.

Finally, we remark that an alternative spin-wave analysis ignores vortices \cite{Buchler12} and predicts a straight line $\alpha_c=3$ for the phase boundary between the XY and CSB phases.
However, the RG equations include the effect of vortices and predict a phase boundary at $3-\alpha_c-1/(2K)=0$; for the perturbative value of $K$ computed in \cite{supp}, we find the dashed line in Fig.~\ref{Fig. SpinHalf} that captures the qualitative trend of the phase boundary near $J_z=0$.

{\it Phase transitions.---}The ferromagnetic (FM) phase for $J_z< -1$ is connected to the CSB and XY phases at $J_z> -1$ via a first-order transition.  The phase transition between the XY and the antiferromagnetic (AFM) phases is the Berezinskii-Kosterlitz-Thouless (BKT) transition, which is well understood for short-range interactions \cite{GiamarchiBook,SachdevBook}.
We are mainly interested in the phase transition from the CSB phase to the XY phase described by Eq.~(\ref{Eq. Action 2}).
Below we derive the full RG flow that goes beyond Eq.~(\ref{Eq. RG for g and gLR}).
\begin{figure}[b]
   \centering
     \includegraphics[width=5.5cm]{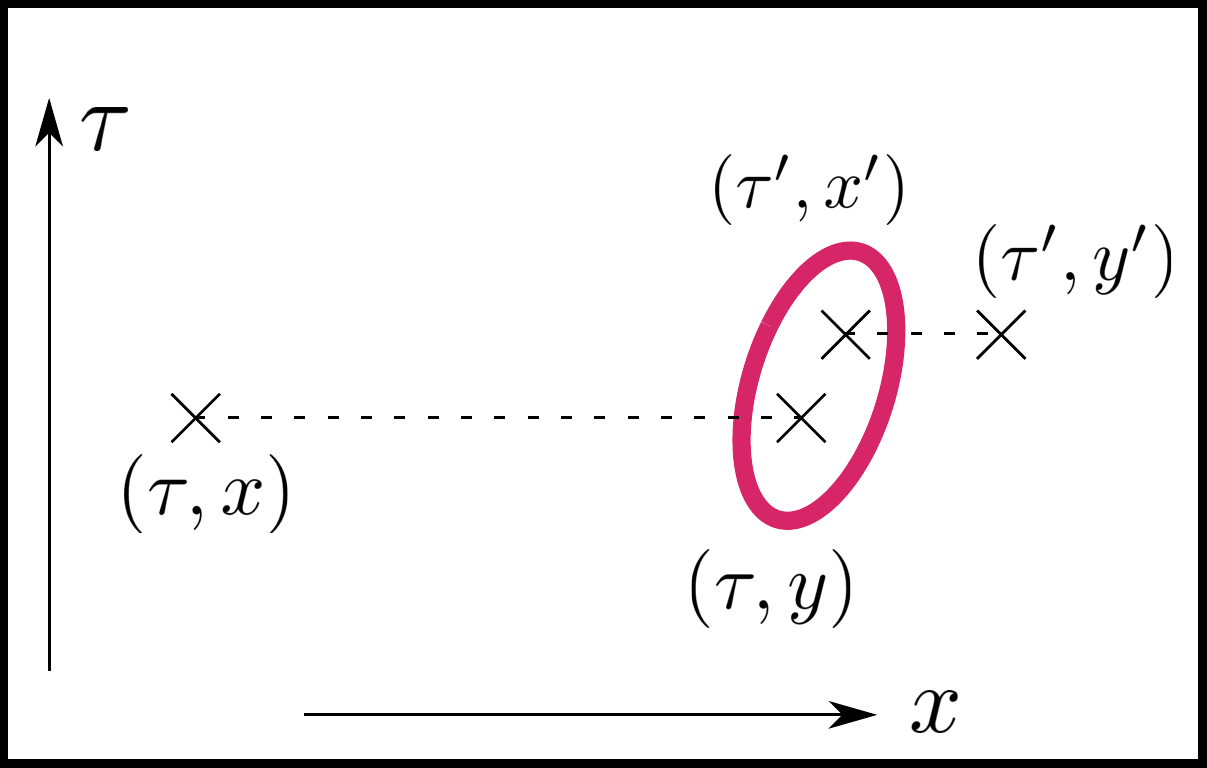}
     \caption{Second-order correction ($\propto g_{\rm LR}^2$) to the RG flow of the long-range interaction term. Each insertion of this term creates two vortices at the same imaginary time $\tau$, but at different spatial coordinates $x$ and $y$. With two such insertions, illustrated in this figure, nearby vortices from different pairs may be neutralized. This will renormalize the original long-range interaction term.}
     \label{Fig. 2ndOrder}
\end{figure}

We first consider the RG flow of the parameter $K$.
Since the interaction term in Eq.~(\ref{Eq. Action 2}) is nonlocal in space but local in time,
we find a renormalization of $(\nabla\theta)^2$, but not $(\partial_\tau \theta)^2$, to  first order in $g_{\rm LR}$.
This implies that $u K$ is renormalized linearly in $g_{\rm LR}$, while $K/u$ is unrenormalized to this order. Note that the velocity $u$ is also renormalized in the absence of an effective Lorentz symmetry, but, to the leading order, we can eliminate its flow via $K/u ={\rm const}+{\cal O}(g_{\rm LR}^2)$.

\begin{figure}[h]
   \centering
    \includegraphics[width=6cm]{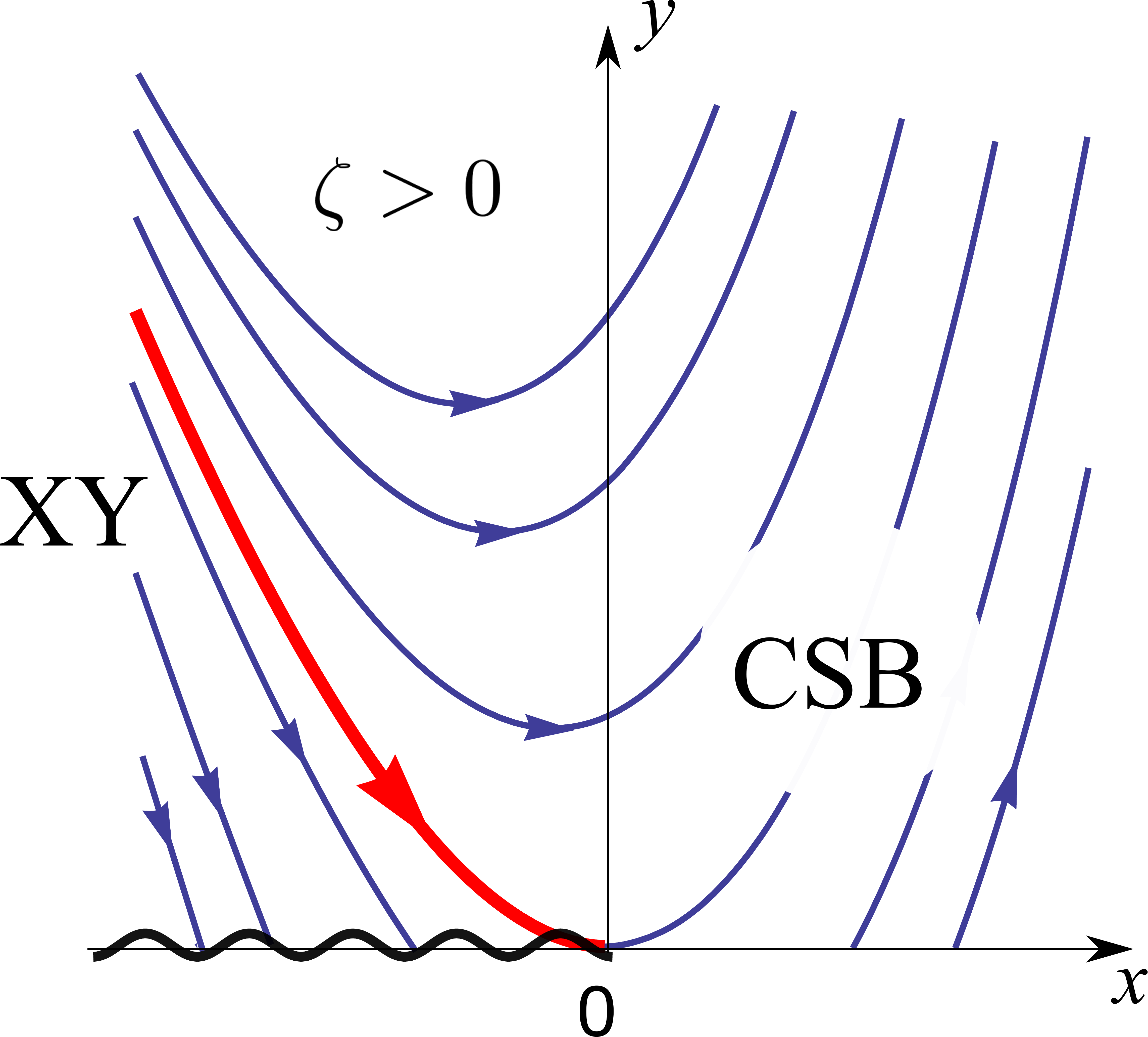}
    \caption{The RG flow in the vicinity of the phase transition, denoted by the thick (red) line, between the CSB phase and the XY phase. Here $x=3-\alpha-1/(2K)$ and $y=g_{\rm LR}$. The RG flow is given by $y\sim x^2+\zeta(1+x)$, where the parameter $\zeta$ quantifies the distance from the critical point. For $x < 0$, the $\zeta=0$ contour describes the critical line. The flows with $\zeta>0$ and those with $\zeta < 0$ and $x > 0$ proceed to infinity characterizing the CSB phase. The trajectories with $\zeta < 0$ and $x<0$ flow to the wavy line characterizing the XY phase.}%
    \label{Fig. RGflow}%
\end{figure}

The RG flow for $g_{\rm LR}$ is given by the second equation in Eq.~(\ref{Eq. RG for g and gLR}); however, it also receives corrections at the quadratic order in $g_{\rm LR}$:
At this order, two vortices  from different insertions of $g_{\rm LR}$ can neutralize each other at close distances, while the remaining vortices form an interaction of the same form as the second line of Eq.~(\ref{Eq. Action 2}), see Fig.~\ref{Fig. 2ndOrder}.
Putting the above considerations together, we find the RG equations to  first nonzero order:
\begin{align}\label{Eq. A and B}
  \frac{dK}{dl} &= A_K\, g_{\rm LR}\,, \nonumber\\
  \frac{dg_{\rm LR}}{dl} &= \left[3-\alpha-1/(2K)\right]g_{\rm LR} + B_K \, g_{\rm LR}^2\,,
\end{align}
with $A_K$ and $B_K$ depending on the parameter $K$, see \cite{supp} for details. In particular, $A_K>0$ since the interaction tends to pin the field, and thus suppresses its fluctuations by increasing $K$.
One can also argue that $B_K>0$ by inspecting the RG equations near the fixed point \cite{supp}.
To find the critical behavior near the fixed point, we expand the above equations in its vicinity by defining $x=3-\alpha-1/(2K)$ and $y=g_{\rm LR}$ for notational convenience. The RG flow equations are then given by
\begin{equation}\label{Eq. x and y}
  \frac{dx}{dl} = A \,y, \qquad \frac{dy}{dl} = x y + B \,y^2\,,
\end{equation}
where $A=A_K/2K^2$ and $B=B_K$ with the substitution $K\to 1/[2(3-\alpha)]$.
The above equations define a new universality class distinct from the usual BKT transition:
The flow equation for $x$ starts at the linear order in $y$ (as opposed to $y^2$), and the correction to the RG equation for $y$ appears at the quadratic order which should be kept (as opposed to $y^3$).
Indeed, the RG flow for the usual BKT transition is unchanged under $y \to -y$ [an example of which is the sine-Gordon model (\ref{Eq. Hamiltonian SR}), where a change of $g\to -g$ can be simply undone by $\phi \to \phi +\pi/4$], but there is no such requirement for long-range interactions, hence the appearance of lower-order terms in Eq.~(\ref{Eq. x and y}).
The corresponding RG flow diagram is shown in Fig.~\ref{Fig. RGflow}.
The flow trajectory near the transition point has a parabolic form given by (with suitably rescaled variables $x$ and $y$)
\begin{equation}
  y\sim x^2 + \zeta (1+x)\,,
\end{equation}
where $\zeta$ parameterizes the distance from the critical trajectory; for $\zeta=0$, one finds the critical trajectory $y\sim x^2$. The RG flow and the form of the critical trajectory are distinctly different from the BKT transition, where the trajectories are hyperbolic, and the critical trajectory is a wedge $y\sim |x|$ rather than a parabola $y\sim x^2$ (for $x<0$) \cite{Cardy96}.

At large distances, the correlation function in the CSB phase approaches a constant [Eq.~(\ref{Eq. cs breaking criterion})]; however, at short distances, the system still exhibits power-law decaying correlations predicted by the XY model. We shall denote the length scale that separates these two regimes by $\xi$.
This length scale diverges near the phase transition as $\zeta \to 0^+$. To find the critical behavior of $\xi$, we solve the RG equation
\(
  {dx}/{dl}=A y \sim x^2 +\zeta(1+x)
\)
 for $x(l)$. However, the RG equation is perturbative, and should not be trusted when $x(l)\sim y(l)\sim 1$ \cite{Cardy96}. This occurs for a value of $l^*\sim 1/\sqrt{\zeta}$, which then determines the scaling of the length scale $\xi$ with $\zeta$ as
\begin{equation}
  \xi \sim e^{l^*}\sim e^{1/\sqrt{\zeta}}\,.
\end{equation}
(A coefficient of order unity is ignored in the exponent.) This relation is reminiscent of the BKT transition where $\zeta$ should be identified as the distance from the critical temperature, and $\xi$ as the correlation length \cite{Cardy96}. In our case, $\zeta$ is simply a parameter that quantifies the distance from the critical trajectory; one can take it, for example, to be the difference of the exponent $\alpha$ from its critical value $\alpha_c$.

{\it Experimental detection.---}Our model Hamiltonian can be realized by optical-dipole-force-induced spin-spin interactions in a trapped ion chain \cite{deng_effective_2005}. For $J_z=0$ and $0.5<\alpha<2$, the dynamics of the Hamiltonian, Eq.\ \eqref{Eq. XXZ Hamiltonian}, has already been simulated experimentally, with measurements available for individual spins \cite{Richerme14,Jurcevic14}. In order to observe the continuous CSB phase and related phase transitions, we can experimentally add a tunable-strength magnetic field in the $x$-$y$ plane.
The ground state for a finite-size system can be adiabatically prepared if we ramp down the magnetic field all the way to zero and slowly enough compared to the energy gap \cite{kim_quantum_2010,islam13}. Then, by measuring the spin correlations, we can confirm the existence of long-range order and of the CSB phase.

{\it Conclusion and outlook.---}In this work, we have considered a 1D spin Hamiltonian with long-range interactions, and shown that a phase with continuous symmetry breaking emerges for sufficiently slowly decaying power-law interaction.
In particular, we have found a new universality class describing the transition from the CSB to the XY phase, similar to, but distinct from, the BKT transition. It is worthwhile exploring continuous symmetry breaking in more complicated spin systems.
More generally, it is desirable to obtain stringent, and model-independent, bounds on how slowly long-range interactions should decay to give rise to spontaneous symmetry breaking in one-dimensional systems at zero temperature. Furthermore, quantum phase transitions from the CSB phase to other 1D quantum phases  are worth exploring. Long-range interactions also arise in the spin-boson model where spins are strongly coupled to a bosonic bath. It is worth investigating if they would lead to similar CSB phases.

\begin{acknowledgments}
We acknowledge useful discussions with M.\ Kardar, M.\ Foss-Feig, L.\ Lepori, G.\ Pupillo, and D.\ Vodola. This work was supported by ARO, AFOSR, NSF PIF, NSF PFC at JQI, and ARL.
\end{acknowledgments}

\newpage

\setcounter{equation}{0}
\renewcommand{\theequation}{S\arabic{equation}}
\setcounter{figure}{0}
\renewcommand{\thefigure}{S\arabic{figure}}

\onecolumngrid
\begin{center}
\Large\textbf{Supplemental Material}% for ``Continuous symmetry breaking and a new universality class in 1D long-range interacting quantum systems''}
\end{center}

\twocolumngrid

In this supplement, we bosonize the short-ranged Hamltonian (Sec.\ \ref{App: Bosonization of SR}), use spin-wave theory to locate the phase boundary between the FM and XY phases (Sec.\ \ref{App: Spin-wave analysis}), and derive the RG flow of the bosonized long-range interacting model [Eq.~(8) of the main text] (Sec.\ \ref{App: RG for LR}).

\section{Bosonization of the short-ranged Hamiltonian}\label{App: Bosonization of SR}
We consider the XXZ model with nearest and  next-nearest neighbor interactions
\begin{align} \label{eq:h12}
  H_{12}=&J_1 \sum_{i}\left(\Sx_i \Sx_{i+1}+\Sy_i \Sy_{i+1}+\Delta\Sz_i \Sz_{i+1}\right) \nonumber \\
  &+J_2\sum_{i}\left(\Sx_i \Sx_{i+2}+\Sy_i \Sy_{i+2}\right)\,.
\end{align}
We shall treat $\Delta$ and $J_2$ perturbatively and bosonize the Hamiltonian. A closely related Hamiltonian defined as $H_{12}'=H_{12}+H'$ with $H'=J_2\sum_i \Sz_i \Sz_{i+2}$ is studied extensively and serves as a textbook example \cite{SGiamarchiBook,SSachdevBook}. The Hamiltonian $H'_{12}$ takes the form given in Eq.~(3) of the main text
 upon bosonization with \cite{SGiamarchiBook,SSachdevBook}
\begin{align}\label{Eq. parameters}
  &uK =J_1 a, \nonumber \\
  &\frac{u}{K}=J_1 a \left[1+\frac{4(\Delta +2J_2/J_1)}{\pi}\right], \\
  & g= a J_1 \Delta-6a J_2\,, \nonumber
\end{align}
where $a$ is the lattice spacing.
Therefore, we just need to bosonize $H'$ and subtract it from Eq.~(3) of the main text
whose parameters are to be substituted from Eq.~(\ref{Eq. parameters}). In this section, we closely follow the steps outlined in Ref.~\cite{SGiamarchiBook}.
Exploiting the Jordan-Wigner transformation, we first cast $H'$ in terms of fermionic operators $c_i$ as
\begin{equation}
  H'=J_2 \sum_i \left( c_i^\dagger c_i-\frac{1}{2}\right)\left( c_{i+2}^\dagger c_{i+2} -\frac{1}{2}\right).
\end{equation}
The operators $c_i$ are mapped to a fermionic field in the continuum as
\begin{equation}
  \psi(x_i)=\frac{1}{\sqrt{a}} \,c_i\,.
\end{equation}
One can decompose the field $\psi$ into the left and right moving modes in the vicinity of the Fermi point, which for the fermion band is simply $k_F=\pi/(2a)$, as
\begin{equation}
  \psi (x_i)= \psi_L(x_i) e^{-i k_F x_{i}}+ \psi_R(x_i) e^{i k_F x_{i}},
\end{equation}
where $\psi_{L/R}$ vary slowly (compared to the lattice spacing). We then have
\begin{widetext}
\begin{align}
  c_i^\dagger c_i-\frac{1}{2} \longrightarrow
  a^2 \!\!\left[\rho_R (x_i)+\rho_L(x_i)+e^{-i2k_F x_i} \psi_R^\dagger(x_i)  \psi_L(x_i)+{\rm h.c.}\right],
\end{align}
where $\rho_{L/R}(x) = \psi^\dagger_{L/R}(x) \psi_{L/R}(x)$ denotes the local density of left/right moving modes. In the bosonization dictionary, we have $\rho \equiv \rho_R+ \rho_L =-\nabla \phi/\pi$, and $\psi_{R/L}(x)\sim \frac{1}{\sqrt{2\pi a_c}}e^{-i\left[\pm \phi(x)-\theta(x)\right]}$ with $a_c$ a short-wavelength cutoff which can be taken to be the same as the lattice spacing $a$. The Hamiltonian can then be written in the continuum as
\begin{align}
  H'=aJ_2 \int dx \, \Big[&\frac{1}{\pi^2}\nabla\phi(x+2a)\nabla\phi(x)+\frac{1}{(2\pi a_c)^2}\left(e^{-i4k_Fa+i 2\left(\phi(x+2a)-\phi(x)\right)}+{\rm h.c.}\right)  \nonumber \\
  & +\frac{1}{(2\pi a_c)^2}\left(e^{-i4k_F x -i k_F a+i 2\left(\phi(r+2a)-\phi(r)\right)}+{\rm h.c.}\right)\Big].
\end{align}
The fact that $e^{i4k_F a}=1$, together with a gradient expansion of the field, yields
\begin{equation}
  H'= aJ_2 \int dx\, \left[-\frac{3}{\pi^2}(\nabla \phi)^2 +\frac{2}{(2\pi a_c)^2} \cos[4\phi(x)] \right].
\end{equation}
\end{widetext}
Comparing the coefficients in this expression against those in Eq.~(\ref{Eq. parameters}), we find that $H_{12}$ maps to Eq.~(3) of the main text
with
\begin{align}\label{Eq. parameters 2}
  &uK =J_1 a, \nonumber \\
  &\frac{u}{K}=J_1 a \left[1+\frac{4\Delta +14J_2/J_1}{\pi}\right], \\
  & g= a J_1 \Delta-5a J_2\,. \nonumber
\end{align}
To directly apply this result to the short-range part (up to the next-nearest neighbor) of the Hamiltonian (1) of the main text,
we first rotate [in Eq.\ (\ref{eq:h12})] every other spin by $180^\circ$ around the $z$-axis, i.e.\ $\Sx_i \to (-1)^i \Sx_i$ and $\Sy_i \to (-1)^i \Sy_i$, to find
\begin{align}
  H_{12}=&J_1 \sum_{i}\left(-\Sx_i \Sx_{i+1}-\Sy_i \Sy_{i+1}+\Delta\Sz_i \Sz_{i+1}\right) \nonumber \\
  &+J_2\sum_{i}\left(\Sx_i \Sx_{i+2}+\Sy_i \Sy_{i+2}\right)\,.
\end{align}
Comparing against Eq.~(1)
with the nearest and next-nearest neighbors, and dropping the term $\Sz_i \Sz_{i+2}$ (proportional to the product of two small parameters), we can now identify $J_1=1$, $J_2=-1/2^\alpha$, and $\Delta=J_z$. Plugging these values in Eq.~(\ref{Eq. parameters 2}), we find
\begin{align}
  K&=\left(1+\frac{4J_z -14/2^\alpha}{\pi}\right)^{-1/2}, \nonumber \\
  u&=a\left(1+\frac{4J_z -14/2^\alpha}{\pi}\right)^{1/2}, \\
  g&=a\,\left(J_z+5/2^{\alpha}\right). \nonumber
\end{align}

\section{Spin-wave analysis near the FM-XY phase boundary}\label{App: Spin-wave analysis}
Consider the Hamiltonian
\[
H=\sum_{i<j}\frac{1}{|i-j|^{\alpha}}(-S_{i}^{x}S_{j}^{x}-S_{i}^{y}S_{j}^{y}+J_{z}S_{i}^{z}S_{j}^{z})\,,
\]
for an arbitrary spin $s$.
For a sufficiently negative $J_{z}$, the ground state is in the
Ising (${\mathbb Z}_2$ degenerate) ferromagnetic phase with all spins fully polarized in the $z$ direction. This state can be regarded as the vacuum state, and spin components can be mapped to bosons, via the
Holstein-Primakoff transformation as $S_{i}^{z}=s-a_{i}^{\dagger}a_{i}$,
$S_{i}^{+}\equiv S_{i}^{x}+iS_{i}^{y}=\sqrt{2s}a_{i}^{\dagger}\sqrt{1-\frac{a_{i}^{\dagger}a_{i}}{2s}}$.
In the low-excitation limit where $\langle a_{i}^{\dagger}a_{i}\rangle\ll s$,
we approximate $S_{i}^{\dagger}\approx\sqrt{2s}a_{i}^{\dagger}$
and map the Hamiltonian $H$ to
\begin{eqnarray}
H & \approx & \sum_{i<j}\frac{s}{|i-j|^{\alpha}}[-J_{z}(a_{i}^{\dagger}a_{i}+a_{j}^{\dagger}a_{j})-a_{i}^{\dagger}a_{j}-a_{j}^{\dagger}a_{i}) \nonumber \\
 & \equiv & -s\sum_{i,j}J_{ij}a_{i}^{\dagger}a_{j}\,,
\end{eqnarray}
where we have ignored the quartic term $a_{i}^{\dagger}a_{i}a_{j}^{\dagger}a_{j}$
since $\langle a_{i}^{\dagger}a_{i}\rangle,\langle a_{j}^{\dagger}a_{j}\rangle\ll s$. Here $J_{ij}=1/|i-j|^{\alpha}$ for $i\ne j$ and $J_{ii}\equiv J_{z}\sum_{j\ne i}J_{ij}$.
The above quadratic Hamiltonian can be brought into a diagonal form in momentum space as $H=\sum_{k}\omega_{k}c_{k}^{\dagger}c_{k}$, with the dispersion
relation in the $N\rightarrow\infty$ limit given by ($q=2\pi k/N$)
\begin{align}
\omega_{q}&=-2s\sum_{r=1}^{\infty}\frac{1}{r^{\alpha}}[J_{z}-\cos(qr)] \nonumber\\
          &=-2s\{J_{z}\zeta(\alpha)+\Re[\text{Li}_{\alpha}(e^{iq})]\}\,,
\end{align}
where $\zeta(\alpha)\equiv\sum_{r=1}^{\infty}1/r^{\alpha}$ is the
Riemann zeta function and $\text{Li}_{\alpha}(e^{iq})$ is
the poly-log function. We note that $\Re[\text{Li}_{\alpha}(e^{iq})]$
has its maximum at $q=0$, at which point it is equal to $\zeta(\alpha)$, for all $\alpha>1$. Therefore,
$\omega_{q}$ has its minimum at $q=0$,
\begin{equation}
\omega_{\rm min}=\omega_{q=0}=-2s(J_{z}+1)\zeta(\alpha)\,.
\end{equation}
Note that for all $\alpha>1$, we have $\omega_{\rm min}>0$ for
$J_{z}<-1$, showing indeed that the system is gapped, and the ferromagnetic
state is the true ground state. Specifically, $\omega_{\rm min}=0$ for $J_{z}=-1$, across which
the system undergoes a first-order phase transition (the ferromagnetic phase remains an exact ground state
of $H$ at $J_{z}=-1$ for any system size). For  $J_{z}>-1$, we find $\omega_{\rm min}<0$, and thus the spin-wave analysis fails.

\section{Renormalization group theory for the long-range interacting model}\label{App: RG for LR}

In this section, we compute the coefficient $A_K$ in the RG flow equation, Eq.~(11),
and outline how the coefficient $B_K$ should be computed. In the process, we will closely follow the methods in Appendix E of Ref.~\cite{SGiamarchiBook}. Let us start with the partition function
\begin{equation}
  Z=\int D\theta \, e^{-I[\theta]},
\end{equation}
with the action $I$ defined in Eq.~(8).
A sharp ultraviolet cutoff $\Lambda$ in momentum space is assumed, that is, the integral in the partition function is over all configurations $\theta_{\bq}$ with $|\bq|<\Lambda$, where we have defined $\bq=(\omega/u,q)$ with $q$ and $\omega$ the momentum and frequency, respectively. We also keep in mind that a cutoff in position space is imposed on long-range interactions as $|x-y|>\lambda$ for some $\lambda$ in the last term of Eq.~(8) of the main text.
Varying the momentum cutoff between $\Lambda$ and $\Lambda'(<\Lambda)$, one can decompose the field $\theta$ into fast and slow modes,
\(
  \theta(\tau, x)= \theta^<(\tau, x)+\theta^>(\tau, x)
\)
with the superscripts $<$ and $>$ corresponding to a sum over Fourier modes with momenta $|\bq|<\Lambda'$ and $\Lambda'<|\bq|<\Lambda$, respectively. These modes simply decouple at the quadratic level of the action $I_0$ as
\(
  I_0=I_0^<+I_0^>\,.
\)
The partition function can be expanded in the powers of the cosine term. One should then integrate out the fast modes to find an effective action in terms of the slow field. To  first order, we find the effective action as
\begin{widetext}
\begin{align} \label{eq:eff}
  I_0^< \,\, -\,\, g_{\rm LR} \int \frac{d\tau dx dy }{|x-y|^\alpha} \cos\left[\theta^<(\tau,x)-\theta^<(\tau,y)\right]
  \, e^{- \int_{\Lambda'<|\bq|<\Lambda} [1-\cos q(x-y)] \frac{\pi  u/K }{\omega^2+u^2q^2}}\,,
\end{align}
where the integral measure in the exponent is $d\omega dq/(2\pi)^2$.
The second term in Eq.\ (\ref{eq:eff}) can be broken into two parts as
\begin{align}\label{Eq. Decomp}
  &-g_{\rm LR} \int \frac{d\tau dx dy }{|x-y|^\alpha} \cos\left[\theta^<(\tau,x)-\theta^<(\tau,y)\right]
  e^{- \int_{\Lambda'<|\bq|<\Lambda} \frac{\pi  u/K }{\omega^2+u^2q^2}}\nonumber \\
  &-g_{\rm LR} \int \frac{d\tau dx dy }{|x-y|^\alpha} \cos\left[\theta^<(\tau,x)-\theta^<(\tau,y)\right]
  e^{- \int_{\Lambda'<|\bq|<\Lambda} [1-\cos q(x-y)] \frac{\pi  u/K }{\omega^2+u^2q^2}}
  \left[1-e^{- \int_{\Lambda'<|\bq|<\Lambda} \cos q(x-y) \frac{\pi  u/K }{\omega^2+u^2q^2}}\right].
\end{align}
The first line in this equation simply renormalizes $g_{\rm LR}$. To see this, note that we first have to rescale space and time coordinates as
\(
\tau=(\Lambda/\Lambda')\tau'
\)
and
\(
x=(\Lambda/\Lambda')x'.
\)
Upon this transformation, the coefficient of the first term is renormalized as
\begin{align}
  g_{\rm LR}(\Lambda')=g_{\rm LR}(\Lambda)\left(\frac{\Lambda}{\Lambda'}\right)^{3-\alpha}e^{- \int_{\Lambda'<|\bq|<\Lambda} \frac{\pi  u/K }{\omega^2+u^2q^2}}=g_{\rm LR}(\Lambda)\left(\frac{\Lambda}{\Lambda'}\right)^{3-\alpha-1/(2K)}.
\end{align}
With the identification $\Lambda'=e^{-dl}\Lambda$, this equation will produce the first term of the flow equation for $g_{\rm LR}$ given by Eq.~(11) of the main text.
To obtain the second term in the latter equation, one should consider the expansion of the cosine term to the quadratic order which we shall discuss later.
Let us now consider the second line in Eq.~(\ref{Eq. Decomp}). First note that the bracket is proportional to $dl$, and thus all the rescaling terms that depend on $\Lambda/\Lambda'$ can be replaced by 1. Furthermore, since the integral over $\bq$ is only for values on the order of the cutoff, $|x-y|$ should be of order  $1/\Lambda$. This suggests that we can expand $\cos\left[\theta^<(\tau,x)-\theta^<(\tau,y)\right] $ in the second line for small values of $|x-y|$. However, one should be careful in expanding this cosine term since the fluctuations of the field are unbounded. This can be remedied by a normal ordering as \cite{SGiamarchiBook}
\begin{align}
  \cos\left[\theta\right] \nonumber =:\cos[\theta]:
   e^{-\frac{1}{2}\left\langle \theta^2\right\rangle},
\end{align}
where the normal-ordered expression can be safely expanded. We find
\begin{align}
  \cos\left[\theta^<(\tau,x)-\theta^<(\tau,y)\right] =\left[1-\frac{(x-y)^2}{2}(\nabla\theta)^2 \right] e^{- \int_{|\bq|<\Lambda'} [1-\cos q(x-y)] \frac{\pi  u/K }{\omega^2+u^2q^2}}.
\end{align}
Putting the above results together, we find a correction to the effective action of the from
\begin{equation}\label{Eq. delta I}
  \delta I = g_{\rm LR} K^{-1} \delta l \int d\tau dX (\partial_X \theta)^2 \int_\lambda^\infty dr\,r^{2-\alpha} e^{-F_\Lambda(r)/2K}J_0(\Lambda r)\,,
\end{equation}
where we have made a change of variables $X=(x+y)/2$ and $r=x-y$, and defined
\begin{equation}
  F_\Lambda(r)=\int_{|\bq|<\Lambda} \frac{d\omega dq}{(2\pi)^2}\,[2-2\cos(q r)]\frac{\pi u}{\omega^2+u^2 q^2}=\int_0^\Lambda \frac{dq}{q}[1-J_0(qr)]\,.
\end{equation}
Also note that $\lambda$ sets a lower bound on the integration over $r$.
Equation (\ref{Eq. delta I}) readily determines the RG flow equations as
\begin{align}
  \frac{d}{dl}\left(K u\right)&=\frac{\pi \, g_{\rm LR}}{K} \int_\lambda^\infty dr\,r^{2-\alpha} e^{-F_\Lambda(r)/2K}J_0(\Lambda r) +{\cal O}(g_{LR}^2)\,, \nonumber \\
  \frac{d}{dl}\left(K/ u\right)&={\cal O}(g_{\rm LR}^2)\,.
\end{align}
The precise form of the RG equations depend on the cutoffs $\Lambda$ and $\lambda$, see App. E of \cite{SGiamarchiBook} for a discussion.
To first order in $g_{\rm LR}$, we have $K/u=c_0=$const and find
\begin{equation}
  \frac{d K}{dl}=g_{\rm LR}\left[\frac{\pi c_0 }{2K^2} \int_\lambda^\infty dr\,r^{2-\alpha} e^{-F_\Lambda(r)/2K}J_0(\Lambda r)\right] +{\cal O}(g_{LR}^2)\,.
\end{equation}
The expression in the bracket gives the value of $A_K$ in Eq.~(11) of the main text.
The short-wavelength cutoff $\lambda$ on long-range interactions can be chosen to be of the order of $\lambda \sim \Lambda^{-1}$.
One can then explicitly see that $A_K>0$.

Derivation of the coefficient $B_K$ in Eq.~(11) of the main text
is rather involved. We shall only outline how one should compute such a term. To this end, we should expand the cosine term in the action (8) of the main text
to  second order in $g_{\rm LR}$. The effective action will then receive a correction of the form (constants of proportionality are ignored)
\begin{align}
  g_{\rm LR}^2\sum_{\epsilon=\pm} \int \frac{d\tau dx dy }{|x-y|^\alpha} \int \frac{d\tau' dx' dy' }{|x'-y'|^\alpha} &\cos\left[\theta^<(\tau,x)-\theta^<(\tau,y)+\epsilon \left(\theta^<(\tau',x')-\theta^<(\tau',y')\right)\right]  \nonumber \\ & \times e^{-\frac{1}{2}\left\langle \left[\theta^>(\tau,x)-\theta^>(\tau,y)+\epsilon \left(\theta^>(\tau',x')-\theta^>(\tau',y')\right)\right]^2\right\rangle},
\end{align}
\end{widetext}
where the expectation value in the exponent should be computed with respect to the quadratic action $I_0^>$.
In a situation where, for example, $(\tau,y)$ and $(\tau',x')$ are nearby points
in space and time (Fig.~2 of the main text), the expectation value finds the nontrivial connected---between terms belonging to different insertions of the interaction term---correlation function $\left\langle \theta^>(\tau,y)\theta^>(\tau',x')\right\rangle$. Also the cosine term in the first line of the above equation, with an appropriate choice of $\epsilon$ and a careful normal ordering, becomes $\cos\left[\theta(\tau,x)-\theta(\tau,y')\right]$; we have used the fact that $\tau\approx \tau'$, while $x$ and $y'$ may be far from each other.
This resembles the interaction term in Eq.~(8) of the main text
up to a multiplicative power law. In fact, the integration over $y$ or $x'$ will generate a series of power laws starting with $1/|x-y'|^\alpha$, which thus generates a term with the same form as the original interaction term. Therefore, the renormalization of $g_{\rm LR}$ finds a correction at the quadratic order in $g_{\rm LR}$.

One can argue that $B_K>0$ by inspecting the RG equations in Eq.~(11) of the manuscript near the fixed point where $3-\alpha-1/(2K)=0$ and $g_{\rm LR}=0$. Going away from this point by increasing $g_{\rm LR}$, the system is expected to be in the ordered phase where the value of $\theta$ is pinned. This implies that $dg_{\rm LR}/dl>0$, which, in turn, requires $B_K>0$.
\\

\end{document}